# Neural Network-based Graph Embedding for Cross-Platform Binary Code Similarity Detection

Xiaojun Xu
Shanghai Jiao Tong University
xuxj@apex.sjtu.edu.cn

Chang Liu
University of California, Berkeley
liuchang@eecs.berkeley.edu

Qian Feng
Samsung Research America
qian.feng1@samsung.com

Heng Yin
University of California, Riverside
heng@cs.ucr.edu

Le Song
Georgia Institute of Technology
lsong@cc.gatech.edu

Dawn Song
University of California, Berkeley
dawnsong@cs.berkeley.edu

## ABSTRACT

The problem of cross-platform binary code similarity detection aims at detecting whether two binary functions coming from different platforms are similar or not. It has many security applications, including plagiarism detection, malware detection, vulnerability search, etc. Existing approaches rely on approximate graph-matching algorithms, which are inevitably slow and sometimes inaccurate, and hard to adapt to a new task. To address these issues, in this work, we propose a novel neural network-based approach to compute the *embedding*, i.e., a numeric vector, based on the control flow graph of each binary function, then the similarity detection can be done efficiently by measuring the distance between the embeddings for two functions. We implement a prototype called Gemini. Our extensive evaluation shows that Gemini outperforms the state-of-the-art approaches by large margins with respect to similarity detection accuracy. Further, Gemini can speed up prior art's embedding generation time by 3 to 4 orders of magnitude and reduce the required training time from more than 1 week down to 30 minutes to 10 hours. Our real world case studies demonstrate that Gemini can identify significantly more vulnerable firmware images than the state-of-the-art, i.e., Genius. Our research showcases a successful application of deep learning on computer security problems.

## 1 INTRODUCTION

Given two binary functions, we would like to detect whether they are similar. This problem is known as "binary code similarity detection," which has many security applications, such as plagiarism detection, malware detection, vulnerability search, etc. Among these security applications, vulnerability search in firmware images of IoT devices is particularly critical and more crucial than ever. A single bug at source code level may spread across hundreds or more devices that have diverse hardware architectures and software platforms. The study by Cui et al. showed that 80.4% of vendor-issued firmware is released with multiple known vulnerabilities, and many recently released firmware updates contain vulnerabilities in third-party libraries that have been known for over eight years [12].

Security practitioners face an increasing need to quickly detect similar functions directly in binaries across multiple platforms, e.g., x86, ARM, or MIPS. Only recently, researchers have started to tackle the problem of cross-platform binary code similarity detection [16, 18, 31]. These efforts propose to extract directly from binary code various robust platform-independent features for each node in the control flow graph to represent a function. Then, to conduct a binary code similarity detection, a graph matching algorithm is used to check whether two functions' control flow graph representations are similar [16, 31]. On the other hand, Genius [18] learns high-level feature representations from the control flow graphs and encodes (i.e., embeds) the graphs into *embeddings* (i.e., high dimensional numerical vectors). To compute the embedding of a binary function, however, it also relies on graph matching algorithms to compute the similarity between the target function and a codebook of binary functions.

Unfortunately, such graph matching-based approaches have two inevitable drawbacks. First, the similarity function approximated by fixed graph matching algorithms is hard to adapt to different applications. For example, given two pieces of binary code which differ in only a few instructions, in the application of plagiarism detection, they may be considered as similar, since the majority of the code is identical; but in the application of vulnerability search, they may be considered dissimilar, since a few instructions' difference may fix an important vulnerability. A manually designed similarity function cannot fit in both scenarios by nature.

Second, the efficiency of all similarity detection approaches based on graph matching is bounded by the efficiency of the graph matching algorithms (such as bipartite graph matching). However, the graph matching algorithms are slow, i.e., requiring super-linear runtime in the graph size. Thus such approaches are inevitably inefficient.

In recent years, deep learning [28] has been applied to many application domains, including binary analysis [42], and has shown stronger results than other approaches. The advantage of deep neural networks is that they can represent a binary analysis task, e.g., generating embedding for a binary function, as a neural network whose parameters can be trained end-to-end, so that it relies on as little domain knowledge (e.g., graph matching in previous approaches) as possible. Further, a deep neural network-based approach can be adaptive by design, since the neural network can be trained with different data to fit into different application scenarios or tasks. Also, a deep neural network model can be computed efficiently, i.e., with runtime linear to the input-size and the network-size.

Inspired by these advantages, in this work, we propose a deep neural network-based approach to generate embeddings for binary functions for similarity detection. In particular, assuming a binary function is represented as a control-flow graph with attributes attached to each node, we use a *graph embedding network* to convert the graph into an embedding. Previously, graph embedding networks have been proposed for classification and regression tasks in domains such as molecule classification [13]. However, our work is in similarity detection, which is different from classification, and thus their approach does not apply to our task directly. Instead, we propose a new approach to computing graph embedding for similarity detection, by combining graph embedding networks into a Siamese network [7] that naturally captures the objective that the graph embeddings of two similar functions should be close to each other and vice versa. This entire network model can then be trained end-to-end for similarity detection.

Further, we design a new training and dataset creation method using a default policy to *pre-train* a task-independent graph embedding network. Our approach constructs a large-scale training dataset using binary functions compiled from the same source code but for different platforms and compiler optimization levels. Our evaluation demonstrates that this task-independent model is more effective and generalize better to unseen functions than the state-of-the-art graph matching-based approach [18].

One advantage of the neural network-based approach is that the pre-trained model can be *retrained* quickly in the presence of additional supervision to adapt to new application scenarios. Our evaluation shows that with such additional supervision, the retrained model can efficiently adapt to novel tasks. Different from previous approaches such as Genius, which would take more than a week to retrain the model, training a neural network is very efficient, and each retraining phase can be done within 30 minutes. This efficiency property enables practical usage of the retraining to improve the quality of similarity detection.

We have implemented a prototype called Gemini. Our evaluations demonstrate that Gemini outperforms the state-of-the-art approaches such as Genius [18] by large margins with respect to both accuracy and efficiency. For accuracy, we apply Gemini to the same tasks used by Genius to evaluate both task-independent and task-specific models. For the former, the AUC (Area Under the Curve) of our pre-trained task-independent model is 0.971, whereas AUC for Genius is 0.913. For the latter, from a real-world dataset, our task-specific models can identify on average 25 more vulnerable firmware images than Genius among top-50 results. Note that previous approaches do not provide the flexibility to incorporate additional task-specific supervision efficiently. Thus the retraining process is a unique advantage of our approach over previous work.

For efficiency, Gemini is more efficient than Genius in terms of both embedding generation time and training time. For embedding generation, Gemini is 2400× to 16000× faster than the Genius approach. For training time, training an effective Gemini model requires less than 30 minutes, while training Genius requires more than one week.

In a broader scope, this work showcases a successful example of how to apply deep learning to solve important and emerging computer security problems and substantially improves over the state-of-the-art results.

We summarize our contributions as follows:

- We propose the first neural network-based approach to generating embeddings for binary functions;
- We propose a novel approach to train the embedding network using a Siamese network so that a pre-trained model can generate embedding to be used for similarity detection;
- We propose a retraining approach so that the pre-trained model can take additional supervision to adapt to specific tasks;
- We implement a prototype called Gemini. Our evaluation demonstrates that on a test set constructed from OpenSSL, Gemini can achieve a higher AUC than both Genius and other state-of-the-art graph matching-based approach;
- Our evaluation shows that Gemini can compute the embedding 3 to 4 orders of magnitude faster than prior art, i.e., Genius;
- We conduct case studies using real-world firmware images. We show that using Gemini we can find significantly more vulnerable firmware images than Genius.

## 2 BINARY CODE SIMILARITY DETECTION

In this section, we first use cross-platform binary code search as an example to explain the problem to design a similarity detection function. We then explain existing approaches, while providing a demonstration on how an efficient embedding function can help with designing a similarity function. In the end, we present our approach to use a neural network as the embedding function, and the benefits of such an approach.

### 2.1 Motivation Problem: Cross-Platform Binary Code Search

Consider the problem of cross-platform binary code similarity detection. Given a binary function of interest (e.g., one that contains the Heartbleed vulnerability), we would like to examine a large corpus of binary functions (e.g., ones extracted from firmware images of various IoT devices) and quickly and accurately identify a list of candidates that are semantically equivalent or similar to the function of interest. We call the binary function of interest *the query function*, and the corpus of binary functions *the target corpus*.

A technique for this problem can be applied to many security applications, such as bug search in firmware images and plagiarism detection in binary code, etc.

The core of this problem is the design of a function to detect whether two functions are similar or not. An effective approach for this problem needs to achieve the following design goals:

- **Binary only.** In practice, we often do not have access to the source code of the binary functions. As a result, an effective similarity detection and code search technique must work on binary code directly.
- **Cross-platform support.** Since the query function and the functions in the target corpus may come from different hardware architectures and software platforms, an effective binary search technique must tolerate the syntactic variations introduced by different platforms and capture the intrinsic characteristics of these binary functions.
- **High precision.** An effective binary code similarity detector should be able to assign a high score to a pair of similiar functions, and a low score to a pair of irrelevant ones.
- **High efficiency.** The similarity function should be computed efficiently for the vulnerability search system and other applications to scale to a large target corpus.
- **Adaptive.** When domain experts can provide similar or dissimilar examples, the similarity function should be able to be adapted quickly to these examples for the domain-specific application.

## 2.2 Existing Techniques

While there has been a series of efforts on binary code matching and search, most of them only work on binary code for a single platform [14, 32].

Only recently, researchers have started to tackle this problem in a cross-platform setting. These efforts propose to extract various features from binary code directly that are robust enough to persist across different architectures and compiler optimization options.

**Pairwise Graph Matching.** The approach by Pewny et al. [31] extracts input–output pairs for each basic block as its feature (or label) in the control flow graph, and then it performs graph matching. Unfortunately, this is a very expensive process: the calculation of input–output pairs and the graph matching are both expensive.

To improve efficiency, discovRE [16] is proposed to extract more lightweight syntax level features (e.g., the number of arithmetic instructions and the number of call instructions) instead to speed up the feature extraction and to apply pre-filtering by simple function-level features before graph matching to improve the search efficiency. However, according to Feng et al. [18], this pre-filtering approach is not reliable and may cause significant degradation in search accuracy. Fundamentally, both approaches rely on pairwise graph matching to detect similarity, which are inevitably inefficient.

**Graph Embedding.** In order to achieve scalability and high accuracy simultaneously, we would like to learn an indexable feature representation from the control flow graph. In other words, we need to encode (i.e., embed) a graph representation into an embedding, i.e., a numeric feature vector. In doing so, the similarity function can be computed as an easy-to-compute distance function between two vectors, which is thus efficient. Also, the feature vectors can be indexed using a locality sensitive hashing (LSH) based database so that a search query can be executed in $O(1)$ time.

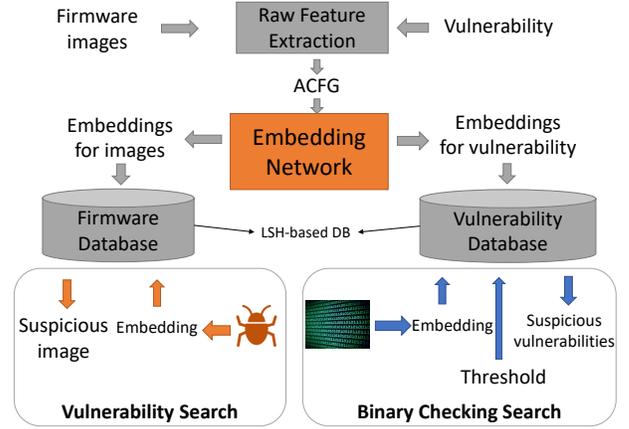

Figure 1: Cross-platform Binary Code Search Workflow

| Type | Attribute name |
|---|---|
| Block-level attributes | String Constants |
| | Numeric Constants |
| | No. of Transfer Instructions |
| | No. of Calls |
| | No. of Instructions |
| | No. of Arithmetic Instructions |
| Inter-block attributes | No. of offspring |
| | Betweenness |

Table 1: Basic-block attributes

Feng et al. [18] were the first to apply this approach to the vulnerability search problem. They proposed Genius, a graph embedding workflow, illustrated in Figure 1. Given a binary function (from either a firmware image or a known vulnerability), Genius first extracts raw features in the form of an *attributed control flow graph (ACFG)*. In an ACFG, each vertex is a basic block labeled with a set of attributes. Table 1 lists six block-level attributes and two inter-block attributes used in Genius. Figure 2 illustrates an ACFG for a function in OpenSSL containing the Heartbleed vulnerability. Each ACFG is then converted into a high-level embedding, which is stored into a hash table using locality sensitive hashing (LSH). Consequently, to identify a set of binary functions that are similar to the query function, we just need to find the corresponding embedding of the query function and find the nearby embeddings in the target corpus.

The key component is how to convert ACFGs into their embeddings. Genius takes a codebook-based approach to embedding an ACFG. It uses a clustering algorithm to train a *codebook* consisting of a number of representative ACFGs identified for each cluster. Then, to convert an ACFG to a feature vector, Genius measures the similarity between the specified ACFG and each representative

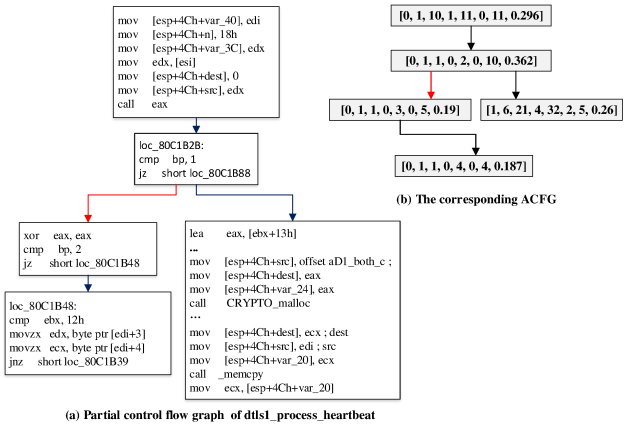

(a) Partial control flow graph of dtls1_process_heartbeat

(b) The corresponding ACFG

**Figure 2: An example of a code graph on Function** `dtls1_process_heartbeat` **(Heartbleed vulnerability)**

ACFG in the codebook using the bipartite graph matching algorithm. Consequently, these similarity measures form the feature vector of the specified ACFG.

While the idea of graph embedding is inspiring and convincing, the use of a codebook and graph matching has several limitations. First, codebook generation is a very expensive process, as pairwise graph matching has to be conducted for each pair of control flow graphs in the training dataset, and then spectral clustering needs to be performed. As a result, the quality of the generated codebook is limited by the scale of the training dataset. Second, the runtime overhead of graph embedding increases linearly with the codebook size (i.e., the number of control flow graphs in the codebook). Therefore, the codebook size has to be small,[1] which confines the fidelity of graph encoding. Last but not least, the search accuracy of this approach is ultimately bounded by the quality of bipartite graph matching [35]. As an approximate algorithm, bipartite graph matching may not always produce optimal matching results.

### 2.3 Neural Network-based Embedding Generation

In this paper, we propose to take a neural network-based approach to embedding an ACFG, to overcome the limitations in previous graph matching-based approaches. Our approach employs a neural network to transform an ACFG into an embedding. We will discuss the details in Section 3. In doing so, our approach has several advantages over previous work:

**Better accuracy.** Our neural network-based embedding can achieve significantly better accuracy than both bipartite graph matching and Genius, for two main reasons. First, neural network-based graph embedding does not rely on bipartite graph matching at all. Instead, it evaluates the graphical representation as a whole, by iteratively propagating embedding throughout the control flow graph. Second, the parameters in the neural network are automatically learned to maximize our embedding objective: the distance between the embeddings of two similar ACFG should be minimized,

---

[1]16 was chosen as the codebook size in the paper by Feng et al. [18]

whereas the distance between the embeddings of two dissimilar ACFGs should be maximized. Further, the neural network-based approach allows the model to be retrained with additional supervision from domain experts to better adapt to a new task/scenario, so as to further improve the accuracy.

**Higher embedding efficiency.** The graph embedding in Genius is very slow, as it has to perform bipartite graph matching with each ACFG in the codebook. In comparison, our neural network model is cheap to compute. Further, all computations in the neural network can be parallelized to leverage the massively parallel computing hardware (i.e., multi-core CPUs and GPUs). Another performance boost comes from not requiring inter-block attributes. In order to achieve good graph matching results, Genius extracts inter-block attributes: the number of offspring and betweenness, which are on average 8× more expensive to compute than the block-level attributes. Our neural network model, on the other hand, requires only basic block-level attributes and the number of offspring (which is cheap to compute) to achieve high accuracy. The neural network model already incorporates the inter-block relation information into the embedding; thus these inter-block attributes (e.g., betweenness) are not needed to achieve high accuracy.

**Faster offline training.** In order to compute a codebook, Genius needs to compute a distance matrix for a large set of training ACFGs, whose time complexity is quadratic in the number of training samples, and linear in the cost of the bipartite graph matching algorithm. In contrast, the neural network approach only requires training for a constant number of epochs, each of which has a time complexity linear in the size of the training data. As a result, Genius requires more than 1 week to generate the codebook, while our approach can train a neural network model within 30 minutes, which enable practical applications using retraining.

## 3 NEURAL NETWORK-BASED MODEL FOR EMBEDDING GENERATION

We first introduce the code similarity embedding problem in Section 3.1 and then present an overview of our solution in Section 3.2. We then explain two important modules of our approach, i.e., the graph embedding network and the overall architecture (Section 3.3), and training approach (Section 3.4). How to acquire a task-independent model through pre-training and task-specific models through re-training are discussed in Section 3.5.

### 3.1 Code Similarity Embedding Problem

As aforementioned, this code similarity measure can be task dependent. We assume there exists an *oracle* $\pi$ determining the code similarity metric for a given task, which is unknown that we would like to learn. Given two binary program functions $f_1, f_2, \pi(f_1, f_2) = 1$ indicates that they are similar; otherwise, $\pi(f_1, f_2) = -1$ indicates that they are dissimilar.

Here, the oracle $\pi$ is specific to each task, and is typically unknown. In certain tasks, a limited number of instances of $\langle f_1, f_2, \pi(f_1, f_2) \rangle$ triple can be observed. For example, the domain experts may be able to provide some ground truth data about the oracle $\pi$.

The objective of code similarity embedding problem is to find a mapping $\phi$ which maps the ACFG of a function $f$ to a vector representation $\mu$. Intuitively, such an embedding should capture enough information for detecting similar functions. That is, given an easy-to-compute similarity function $Sim(\cdot, \cdot)$, (e.g., cosine function of two vectors), and two program binary functions $f_1, f_2$, $Sim(\phi(f_1), \phi(f_2))$ is large if $\pi(f_1, f_2) = -1$, and is small otherwise.

One advantage of learning the embedding (i.e., the mapping $\phi$) is that it enables efficient computation. The similarity between two functions can be computed using an inexpensive similarity function between two vectors, without incurring the cost of expensive graph matching algorithms.

As aforementioned, using a neural network to approximate the embedding function is particularly appealing, since it can be quickly retrained to adapt to a given task easily when limited task-dependent ground-truth data is provided. Also, computing a neural network-based embedding does not rely on any expensive graph matching algorithms, and thus can be implemented efficiently.

### 3.2 Solution Overview

In this section, we present the key ideas of our solution to the code similarity embedding problem. In this work, we assume the binary code of a function $f$ is represented by its ACFG $g$. In the following, we will use the terms "the binary code [of a function]" and "ACFG" interchangeably.

We design the embedding mapping $\phi$ as a neural network. Since the input is an ACFG, we will leverage previous graph embedding networks from the machine learning community to address the problem [13]. However, in Dai et al.'s work [13], the graph embedding network is designed for a classification problem, which requires label information to train the model.

In contrast, our code similarity embedding problem is not a classification problem. Thus existing approaches do not apply directly, and we need to design a novel approach to train the graph embedding network for the similarity detection problem.

To tackle this challenge, we propose a new learning approach. The idea is that instead of training the graph embedding network $\phi$ to do well on a particular predictive task, we will train $\phi$ to do well on *differentiating the similarity* between two input ACFGs. In particular, we design a Siamese architecture [7] and embed the graph embedding network Structure2vec [13] into it. A Siamese architecture takes two functions as its input, and produces the similarity score as the output. This enables the model to be trained end-to-end with only supervision on a graph-pair $g_1, g_2$ as input and the ground truth $\pi(f_1, f_2)$ as output, without any additional hand-crafted heuristics on how the embeddings should be generated. Thus, such an approach is more robust and easier to adapt to different tasks. We explain more details on this overall architecture and training in Section 3.4.

Training a Siamese architecture requires a large number of pairs of similar functions, as well as pairs of dissimilar functions. However, in most tasks, the ground truth data is limited. To address this issue, we use a *default* policy that considers *equivalent* functions (i.e., binary functions compiled from the same source code) are similar, and *inequivalent* functions are not, so that we can easily generate a large training set given a collection of source code. We can use this dataset to *pre-train a task-independent model*, that can be effective for most tasks. Further, to incorporate the little available ground truth data for a task-specific policy, our approach allows the model to be *retrained* to incorporate *task-specific data*. We explain task-independent pre-training and task-specific re-training in more detail in Section 3.5.

### 3.3 Graph Embedding Network

Our graph embedding network is adapted from Structure2vec, by Dai et al. [13]. Denote an ACFG as $g = \langle \mathcal{V}, \mathcal{E} \rangle$ where $\mathcal{V}$ and $\mathcal{E}$ are the sets of vertices and edges respectively; furthermore, each vertex $v$ in the graph may have additional features $x_v$ which correspond to basic-block features in an ACFG. The graph embedding network will first compute a $p$ dimensional feature $\mu_v$ for each vertex $v \in \mathcal{V}$, and then the embedding vector $\mu_g$ of $g$ will be computed as an aggregation of these vertex embeddings. That is $\mu_g := A_{v \in \mathcal{V}}(\mu_v)$, where $A$ is an *aggregation function*, i.e., summation or average. In this work, we choose $\mu_g = \sum_{v \in \mathcal{V}}(\mu_v)$ and leave the exploration of using other aggregation functions as future work.

In the following, we first explain more details about the generic graph embedding network, and then present the variants instantiated specifically for our ACFG embedding problem.

**Basic Structure2vec Approach.** Structure2vec is inspired by graphical model inference algorithms where vertex-specific features $x_v$ are aggregated recursively according to graph topology $g$. After a few steps of recursion, the network will produce a new feature representation (or embedding) for each vertex which takes into account both graph characteristics and long-range interaction between vertex features. More specifically, we denote $\mathcal{N}(v)$ as the set of neighbors of vertex $v$ in graph $g$. Then one variant of the Structure2vec network will initialize the embedding $\mu_v^{(0)}$ at each vertex as 0, and update the embeddings at each iteration as

$$\mu_v^{(t+1)} = \mathcal{F}(x_v, \sum_{u \in \mathcal{N}(v)} \mu_u^{(t)}), \forall v \in \mathcal{V}. \tag{1}$$

In this fixed-point update formula, $\mathcal{F}$ is a generic nonlinear mapping which we will specify our choice later. Based on the update formula, one can see that the embedding update process is carried out based on the graph topology, and in a synchronous fashion. A new round of embedding sweeping across the vertices will start only after the embedding update for all vertices from the previous round has finished. It is easy to see that the update also defines a process where the vertex features $x_v$ are propagated to the other vertices via the nonlinear propagation function $\mathcal{F}$. Furthermore, the more iterations one carries out the update, the farther away a vertex feature will propagate to distant vertices and get aggregated nonlinearly at distant vertices. In the end, if one terminates the update process after $T$ iterations, each vertex embedding $\mu_v^{(T)}$ will contain information about its $T$-hop neighborhood determined by both graph topology and the involved vertex features.

Instead of manually specifying the parameters in the nonlinear mapping $\mathcal{F}$, we learn these parameters. To train a Structure2vec model which is originally designed for a classification problem, previous work requires a ground truth label for every input graph $g$ to indicate which "class" it belongs to. Then the model is linked with a Softmax-layer, so that the entire model can be trained end-to-end

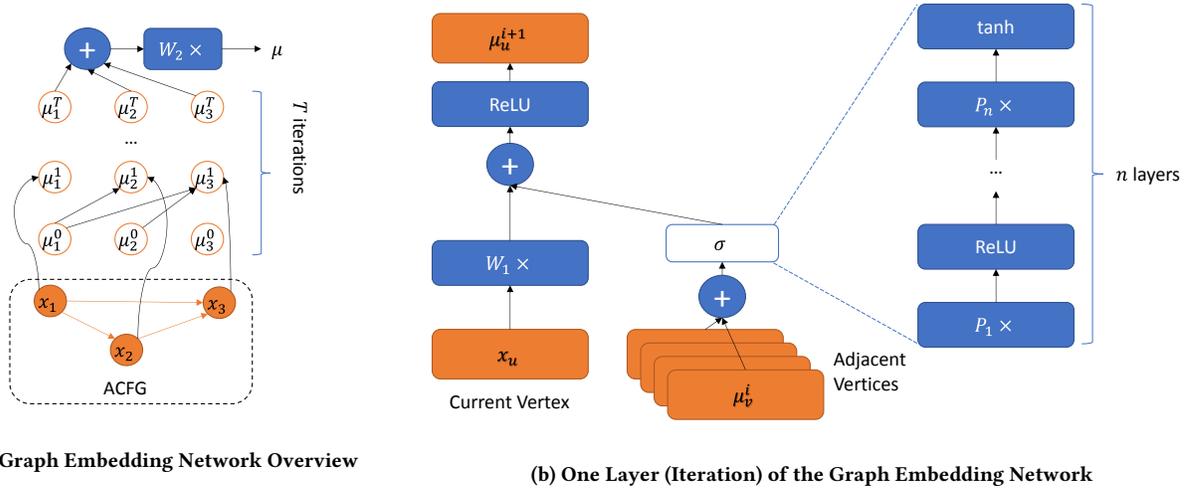

(a) Graph Embedding Network Overview

(b) One Layer (Iteration) of the Graph Embedding Network

Figure 3: Graph Embedding Network

by minimizing the cross-entropy loss. As discussed in Section 3.2, this approach is not applicable to our case since our problem is not a classification problem.

Instead, we train these parameters in $\mathcal{F}$ together with other parameters end-to-end in the Siamese architecture which uses these embeddings for computing similarity, as explained in Section 3.4.

**Our Parameterization for $\mathcal{F}$.** We now discuss our paramemeterization of $\mathcal{F}$ using a neural network. Figure 3 visualizes our network architecture. In particular, we design $\mathcal{F}$ to have the following form

$$\mathcal{F}(x_v, \sum_{u \in \mathcal{N}(v)} \mu_u) = \tanh(W_1 x_v + \sigma(\sum_{u \in \mathcal{N}(v)} \mu_u)) \quad (2)$$

where $x_v$ is a $d$-dimensional vector for graph node (or basic-block) level features, $W_1$ is a $d \times p$ matrix, and $p$ is the embedding size as explained above. To make the nonlinear transformation $\sigma(\cdot)$ more powerful, we will define $\sigma$ itself as an $n$ layer fully-connected neural network:

$$\sigma(l) = \underbrace{P_1 \times \text{ReLU}(P_2 \times ... \text{ReLU}(P_n l))}_{n \text{ levels}}$$

where $P_i$ ($i = 1, ..., n$) is a $p \times p$ matrix. We refer to $n$ as the *embedding depth*. Here, ReLU is the rectified linear unit, i.e., ReLU$(x) = \max\{0, x\}$.

Our novel parameterization of the update function $\mathcal{F}$ together with the iterative update scheme described in Section 3.5 completes our embedding network for ACFGs. The overall algorithm for generating the embedding for each ACFG is summarized in Algorithm 1. In the algorithm, $W_2$ is another $p \times p$ matrix to transform the embedding vector. We denote its output as $\phi(g)$.

### 3.4 Learning Parameters Using Siamese Architecture

In this section, we explain our design of the overall network architecture to train a graph embedding for similarity detection. In particular, we use the Siamese architecture combined with the graph embedding Structure2vec network. The Siamese architecture uses two identical graph embedding networks, i.e., Structure2vec, which join at the top. Each graph embedding network will take one ACFG $g_i$ ($i = 1, 2$) as its input and outputs the embedding $\phi(g_i)$. The final outputs of the Siamese architecture is the cosine distance of the two embeddings. Further, the two embedding networks share the same set of parameters; thus during training the two networks remain identical. The overall architecture is illustrated in Figure 4.

**Algorithm 1** Graph embedding generation
1: **Input:** ACFG $g = \langle \mathcal{V}, \mathcal{E}, \overline{x} \rangle$
2: Initialize $\mu_v^{(0)} = \overline{0}$, for all $v \in \mathcal{V}$
3: **for** $t = 1$ **to** $T$ **do**
4:    **for** $v \in \mathcal{V}$ **do**
5:       $l_v = \sum_{u \in \mathcal{N}(v)} \mu_u^{(t-1)}$
6:       $\mu_v^{(t)} = \tanh(W_1 x_v + \sigma(l_v))$
7:    **end for**
8: **end for**{fixed point equation update}
9: **return** $\phi(g) := W_2(\sum_{v \in \mathcal{V}} \mu_v^{(T)})$

Given a set of $K$ pairs of ACFGs $\langle g_i, g_i' \rangle$, with ground truth pairing information $y_i \in \{+1, -1\}$, where $y_i = +1$ indicates that $g_i$ and $g_i'$ are similar, i.e. $\pi(g_i, g_i') = 1$, or $y_i = -1$ otherwise. We define the Siamese network output for each pair as

$$\text{Sim}(g, g') = \cos(\phi(g), \phi(g')) = \frac{\langle \phi(g), \phi(g') \rangle}{||\phi(g)|| \cdot ||\phi(g')||}$$

where $\phi(g)$ is produced by Algorithm 1.

Then to train the the model parameters $W_1, P_1, \ldots, P_n$, and $W_2$, we will optimize the following objective function

$$\min_{W_1, P_1, \ldots, P_n, W_2} \sum_{i=1}^{K} (\text{Sim}(g_i, g_i') - y_i)^2. \quad (3)$$

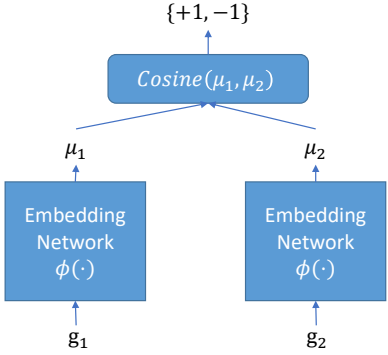

Figure 4: Siamese Architecture

We can optimize the objective (3) with stochastic gradient descent. The gradients of the parameters are calculated recursively according to the graph topology. In the end, once the Siamese network can achieve a good performance (e.g., using AUC as the measure), the training process terminates, and the trained graph embedding network can convert an input graph to an effective embedding suiteable for similarity detection.

### 3.5 Task-independent Pre-training and Task-specific Re-training

Training the model requires a large amount of data on the ground truth about oracle $\pi$, which may be difficult to obtain. To tackle this issue, we construct a training dataset using a *default* policy. This dataset can be used to pre-train a *task-independent* model that is effective for most common tasks. When additional task-specific data becomes available, we allow the pre-trained model to be re-trained quickly to acquire a *task-specific* model. We explain these two approaches below.

**Task-independent Pre-training.** To pre-train a model applicable to most common tasks, intuitively, the generated embedding of each function should try to capture invariant features of the function across different architectures and compilers. We implement this intuition by constructing a dataset as follows using a *default oracle*. Assuming a set of source code is collected, we can compile them into program binaries for different architectures, using different compilers, and with different optimizations. In doing so, the default oracle determines that two binary functions are similar if they are compiled from the same source code, or dissimilar otherwise. To construct the training dataset, for each binary function $g$, one other similar function $g_1$ and one dissimilar function $g_2$ are sampled to construct two training samples, $\langle g, g_1, +1 \rangle$ and $\langle g, g_2, -1 \rangle$. In our evaluation (Section 4.2), we demonstrate that the model pre-trained using this training method performs better than the state-of-the-art graph matching-based approach [18] using the same task evaluated in [18].

**Task-specific Re-training.** Sometimes, the policy used by a specific task may deviate from the default policy used to pre-train the model. In this case, we need an efficient way to fine-tune the learned parameters in the graph embedding network by using a small number of additional data $\langle f, f', \pi(f, f') \rangle$ provided by domain experts. The re-training procedure refines the graph embedding network by incorporating the small number of additional data about the task-specific policy provided by domain experts.

More specifically, assume we get a list of function pairs $\langle g_i, g_i' \rangle$, and their ground truth label $\pi(g_i, g_i')$ from human experts, we can generate additional ACFG pairs to retrain the graph embedding network $\phi(g)$. In particular, for each ACFG pair $g_i, g_i'$ in the provided list, we augment the training set with a pair $\langle g_i, g_i' \rangle$ with pairing information $y_i = \pi(g_i, g_i')$ which is the label from human experts.

Using this augmented dataset, we further train the graph embedding network for a few more (e.g., 5) epochs. In each epoch, the newly added pairs will be sampled more often than the old data (e.g., 50 times more often). After the augmented training is finished, the re-trained network $\phi(\cdot)$ will be deployed for the similarity detection task. Such a retraining procedure allows human experts to provide feedback to the system and the model to be fine-tuned to be adaptive to the task-specific oracle $\pi$, and thus improves the similarity detection accuracy further.

## 4 EVALUATION

In this section, we evaluate Gemini with respect to its search accuracy and computation efficiency. In particular, we evaluate the accuracy of our task-independent pre-trained model using a dataset containing ground truth. We further use real-world datasets to study how well our model can be retrained to adapt to new tasks. In all evaluations, our approach exhibits superior advantages over the state-of-the-art approach [18].[2]

### 4.1 Implementation and Setup

Our system consists of two main components: ACFG extractor, neural network model for graph embedding. We obtain the ACFG extractor, a plug-in to the disassembly tool IDA Pro [1], from the authors of Genius [18], so we can make sure that the raw features extracted from the binary code are consistent with those extracted by Genius. We implement the neural network model in Tensor-Flow [2] in Python.

Our experiments are conducted on a server equipped with two Intel Xeon E5-2620v4 CPUs (32 cores in total) running at 2.1GHz, 96 GB memory, 1TB SSD, and 8 GeForce GTX 1080 GPU cards. During both training and evaluation, only 1 GPU card was used.

**Baseline.** There have been several previous works addressing the bug-search problem: discovRE [16], Multi-HM and Multi-k-HM [31], a centroid-based search [10], and Genius [18]. Feng et al. have demonstrated that the Genius approach is both more accurate and efficient than the other approaches [18]. Therefore, in our evaluation, we consider two baseline approaches evaluated in [18].

- **Bipartite Graph Matching (BGM).** Given two binary functions, we directly compute their similarity score on their ACFGs using the bipartite graph matching as described in Genius. This approach provides a baseline to evaluate the accuracy of pairwise graph matching approaches.
- **Codebook-based Graph Embedding (Genius).** This approach provides a baseline for graph embedding. We contact the authors

---
[2] The code of training and evaluating the pre-trained model is available at https://github.com/xiaojunxu/dnn-binary-code-similarity.

|       | Training | Validation | Testing |
|-------|----------|------------|---------|
| x86   | 30,994   | 3,868      | 3,973   |
| MIPS  | 41,477   | 5,181      | 5,209   |
| ARM   | 30,892   | 3,805      | 3,966   |
| Total | 103,363  | 12,854     | 13,148  |

Table 2: The number of ACFGs in Dataset I

of Genius [18], and obtain the ACFG extraction code and the codebook used in their evaluation. We further implement the codebook generation and embedding generation by ourselves for comparison.

**Datasets.** In our evaluation, we collect four datasets: (1) Dataset I for training the neural network and evaluating the accuracy of the pre-trained model; (2) Dataset II for evaluating the performance of the task-specific model; (3) Dataset III for efficiency evaluation; and (4) a vulnerability dataset (Dataset IV) for case studies.

- **Dataset I:** This dataset is used for neural network training and baseline comparison. It consists of binaries compiled from source code, so that we have ground truth. That is, we consider two ACFGs compiled from the same source code function to be similar, and those from different functions to be dissimilar. In particular, we compile OpenSSL (version 1.0.1f and 1.0.1u) using GCC v5.4. The compiler is set to emit code in x86, MIPS, and ARM, with optimization levels O0-O3. In total, we obtain 18,269 binary files containing 129,365 ACFGs. We split Dataset I into three disjoint subsets of functions for training, validation, and testing respectively. The statistics are presented in Table 2. During the split, we guarantee that no two binary functions compiled from the same source function are separated into two different sets among training, validation and testing sets. In doing so, we can examine whether the pre-trained model can generalize to unseen functions.
- **Dataset II:** We contact the authors of Genius [18] to get the same large-scale dataset used in their paper [18] (referred to as Dataset III in their paper), which includes 33,045 firmware images. Among these images, 8,128 images can be successfully unpacked. These images are from 26 different vendors and for different products such as IP cameras, routers, access points, etc.
- **Dataset III:** To evaluate the efficiency, we construct a dataset with ACFGs of various sizes (i.e., number of vertices in a graph). In particular, we first randomly select 16 firmware images from Dataset II. From these 16 firmware images, there are 82,100 ACFGs whose sizes range from 1 to 1,306. These ACFGs are grouped into sets so that all ACFGs in the same set are the same in size. For any set containing more than 20 ACFGs, we randomly select 20 ones from the set and remove all other ACFGs. In the end, we obtain 3,037 ACFGs in this dataset.
- **Dataset IV:** This dataset contains vulnerable functions obtained from the vulnerability dataset in [18]. In total, it contains of 154 vulnerable functions.

**Training details.** Our neural network model is first pre-trained using Dataset I as below. We use the Adam optimization algorithm [27] and set the learning rate to be 0.0001. We train the Siamese model for 100 epochs. At each epoch, we first construct the training data used for this epoch as follows: for each ACFG $g$ in the training set, we randomly select one ACFG $g_1$ from the set of all ACFGs compiled from the same source function as $g$, and one ACFG $g_2$ from the set of all other ACFGs in the training set. Then we generate two training samples: $\langle g, g_1 \rangle$ with ground truth label +1 and $\langle g, g_2 \rangle$ with label −1. Notice that since we randomly select $g_1$ and $g_2$ for each $g$ independently at each epoch, the training data often vary at different epochs. After the training data is generated for each epoch, it is randomly shuffled before being fed to the training process. Each mini-batch contained 10 ACFG pairs. After every epoch, we measure the loss and AUC on the validation set. During the 100 training epochs, we save the model that achieved the best AUC on the validation set.

By default, the embedding size $p$ is 64 and the embedding depth $n$ is 2. The model runs for $T = 5$ iterations. The basic-block attributes include block-level attributes and the number of offspring, i.e., 7 attributes in total.

### 4.2 Accuracy

In this section, we evaluate the accuracy of the pre-trained model in Gemini. To this end, we construct a similarity testing dataset as follows: from the testing set in Dataset I, for each ACFG $g$ in the set, we randomly select two ACFGs $g_1, g_2$ from the testing dataset, such that the ground truth labels of $\langle g, g_1 \rangle$ and $\langle g, g_2 \rangle$ are +1 and −1 (i.e., from the same source function vs. not) respectively. This similarity testing dataset consists of 26,265 pairs of ACFGs. Note that the testing set is constructed so that no two binary functions compiled from the same source appear in both the training set and the test set. In doing so, we are able to examine the performance of Gemini on unseen functions. Figure 5a illustrates the ROC curves for our neural network model (Gemini) as well as two baseline approaches. We can see that that Gemini outperforms both BGM and Genius by a large margin.

To further examine the performance of Gemini on graphs with different sizes, we split the similarity-accuracy testing set into a large-graph subset and a small-graph subset. The large-graph subset contains only pairs of two ACFGs which both have at least 10 vertices. The small-graph subset contains the rest. The ROC curves of different approaches evaluated over the large-graph subset and the small-graph subset are plotted in Figure 5b and Figure 5c respectively. From both figures, we have consistent observations: 1) Gemini outperforms both BGM and Genius significantly; and 2) Genius outperforms BGM on small graphs, the BGM performs better than Genius on large graphs, and Gemini outperforms both BGM and Genius on large as well as small graphs.

### 4.3 Hyperparameters

In this section, we evaluate the effectiveness of hyperparameters in the Gemini model. In particular, we examine the impact of the number of training epochs, embedding depth, embedding size, ACFG attributes, and number of iterations. The examination of impact of the number of training epochs is using the similarity validation set. We examine other hyperparameters using the similarity testing set. On the entire similiarty testing set, however, the AUC values are almost identical. Since we are more interested in the performance of our model on large graphs, the examinations of other hyperparameters are using the large-graph subset of the similarity test set.

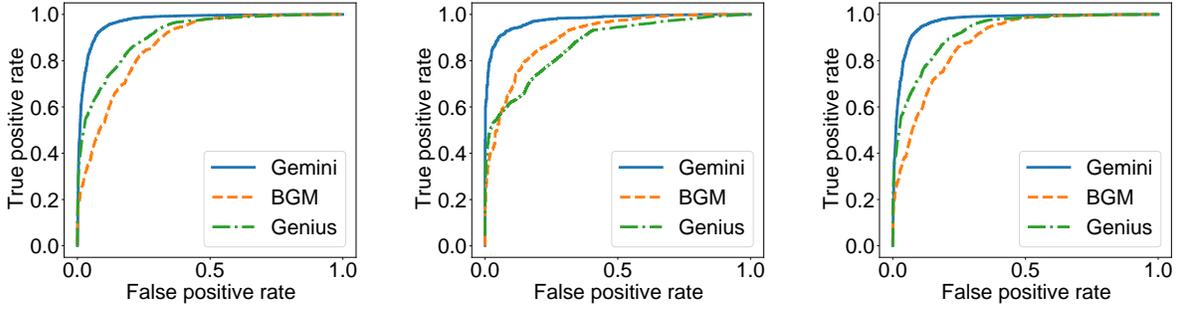

(a) Results on the similarity testing set  (b) Results on the large-graph subset  (c) Results on the small-graph subset

Figure 5: ROC curves for different approaches evaluated on the testing similarity dataset.

**Number of epochs.** We train the model for 175 epochs and evaluate the model over the validation set every 5 epochs for the loss and the AUC. The results are plotted in Figure 7a and Figure 7b. From the figures, we can observe that the loss drops to a low value after 5 training epochs, and then almost remains the same. The lowest loss appears after the model is trained for 100 epochs. We make a similar observation for AUC values, although the highest AUC value appears after the model is trained for 160 epochs. Therefore, we conclude that the model can be quickly trained to achieve reasonably good performance (after 5 epochs).

**Embedding depth.** We vary the number of layers in function $\sigma$ in the Gemini model. From Figure 7c, we observe that when embedding depth is 2, the ROC curve has the largest AUC value. Notice that the original Structure2vec [13] can be viewed as choosing the embedding depth to be 1. We can observe a clear improvement by increasing one more non-linear mapping layer to $\sigma$. However, adding more layers does not help much.

**Embedding size.** In Figure 7d, we can observe that the embedding size to achieve the outter-most ROC curve is 512. However, all curves corresponding to the embedding sizes no smaller than 64 are close to each other. Since a large embedding size requires both longer training time and longer evaluation time, choosing the embedding size to be 64 is a good trade-off between the performance and efficiency. It is worth to note that even when we choose the embedding size to be 16, Gemini is still more effective than both Genius (whose embedding dimensionality is also 16) and BGM.

**ACFG attributes.** We evaluate the accuracy using three different ways to extract the attributes to construct ACFGs. In particular, we consider the attributes to include (1) 6 block-level attributes only (Block); (2) 6 block-level attributes plus the number of offspring (Block+O); and (3) all 8 attributes (Block+O+B). From Figure 7e, we observe that Block+O (7 attributes in total) achieves the best performance. This is unexpected, since a model using the all attributes (Block+O+B) should be more expressive than the model using Block+O. We consider this as an overfitting scenario to the training data. That is, the additional betweeness attribute misleads the model when computing the embedding.

**Number of iterations.** From Figure 7f, we observe that the model achieves the best performance when the number of iterations $T$ is 5 or larger. This is reasonable, since in this dataset all graphs have a size larger than 10. It needs 5-hops to propagate local information on one vertex to most part of the graph.

### 4.4 Efficiency

We evaluate the efficiency of Gemini and Genius for embedding generation using Dataset III. In particular, we measure the latency for the following three tasks: (1) The ACFG extraction time for one function; (2) The embedding generation time from an ACFG; and (3) the overall latency for embedding generation (which includes Task 1 and Task 2). Note that we exclude the disassembly time using IDA pro, since it is the same for both approaches. It usually takes on the order of seconds to disassemble a binary file, which can be amortized over all functions in the binary file.

For embedding generation, we implement several versions for Gemini and Genius respectively. We implement both the CPU and GPU version of Gemini in Tensorflow. In doing so, we can maximally leverage the multi-core hardware to accelerate the performance. For Genius, we implement a single-threaded version as well as a multi-threaded version. Since Genius computes the embedding of a ACFG as the similarity scores between this ACFG and each graph in the codebook using bipartite graph matching, this can be naturally parallelized such that each thread processes one graph in the codebook. The multi-threaded version parallelizes these computations.

**ACFG extraction time.** Figure 6a (one point for each sample) and Figure 6b (average extraction time by different ACFG sizes) illustrate the results. We can observe that extracting only 6 basic-block attributes and extracting 6 basic-block attributes along with the number-of-offspring attribute require similar time, but if we additionally extract the betweenness attribute, it takes on average 8× more time. From Figure 6b, we can observe that the extraction time in general increases along with the ACFG size, but the variation is large. Notice that Genius requires all 8 attributes to achieve its best performance. In contrast, Gemini can aggregate the graph structural information through iterations of embedding updates, and thus can achieve the best performance without the costly betweenness-attribute computation. Therefore, Gemini can improve upon Genius by 8× on average on ACFG extraction.

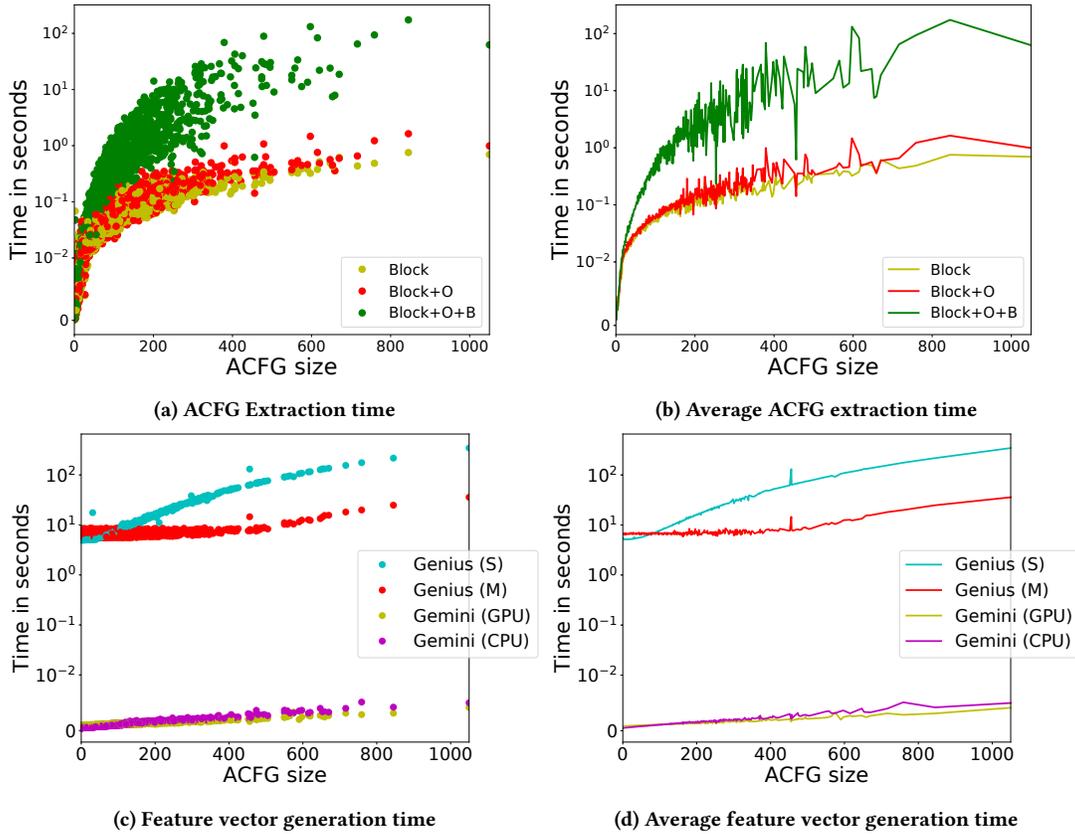

Figure 6: Efficiency evaluation on Dataset III. Figure 6a and Figure 6c plot one point for each sample in Dataset III. In Figure 6b and Figure 6d, we average the running time of all data points with the same ACFG size. Therefore, in these two figures, we have one data point for each ACFG size. In Figure 6a and Figure 6b, "Block" indicates the extraction time of 6 block-level attributes; "+O" indicates that the extraction further includes the number-of-offspring attribute; and "+B" indicates the extraction further includes the betweenness attribute. In Figure 6c and Figure 6d, Gemini (CPU) and Gemini (GPU) denote the CPU and GPU implementations of the Gemini approach respectively. Genius (S) and Genius (M) denote the single-threaded and multi-threaded implementation of Genius respectively.

**Embedding generation time.** Embedding generation time is presented in Figure 6c and Figure 6d. We can observe that the CPU implementation of Gemini runs 2400× to 16000× faster than the multi-threaded version of Genius. On average, the speedup can be as high as 7000×. We attribute this to several reasons. First, the Gemini approach avoids the expensive graph matching algorithm and reduces the time complexity to the number of edges in the graph. Since ACFGs are sparse graphs, i.e., each vertex's out-degree is at most 2, the computation cost is almost linear to the graph size. Second, most computations in the Gemini method can be implemented as matrix operations: matrix multiplication, matrix summation, and element-wise operations over a matrix. All these operations can be parallelized to utilize the underlying multi-core CPUs to achieve speedups with respect to the number of cores. On the other hand, the graph matching algorithm in Genius cannot be easily parallelized. The only speedup comes from processing each element in the codebook in parallel, and thus this speedup is bounded by the number of elements in the codebook. Later, our analysis will show that it is hard to achieve the upper bound of this theoretical speedup in Genius, i.e., the codebook size.

Now we compare the single-threaded version and multi-threaded version of Genius and show that it is hard to achieve the theoretical upper bound of the speedup, i.e., the codebook size. Although the multi-threaded version can run up to 10× faster, the average speedup is just 35%. One reason is that when the codebook used in Genius contains a large graph with more than 500 vertices, the time to process this element dominates the overall feature generation time when the ACFG being processed is small. Further, synchronization of multi-threading introduces additional overhead. Figure 6c and Figure 6d support this observation: (1) when the ACFG is small, the multi-threaded version is similar or even slower than the single-threaded version; and (2) when the ACFG becomes larger, the speedup of multi-threaded also increases.

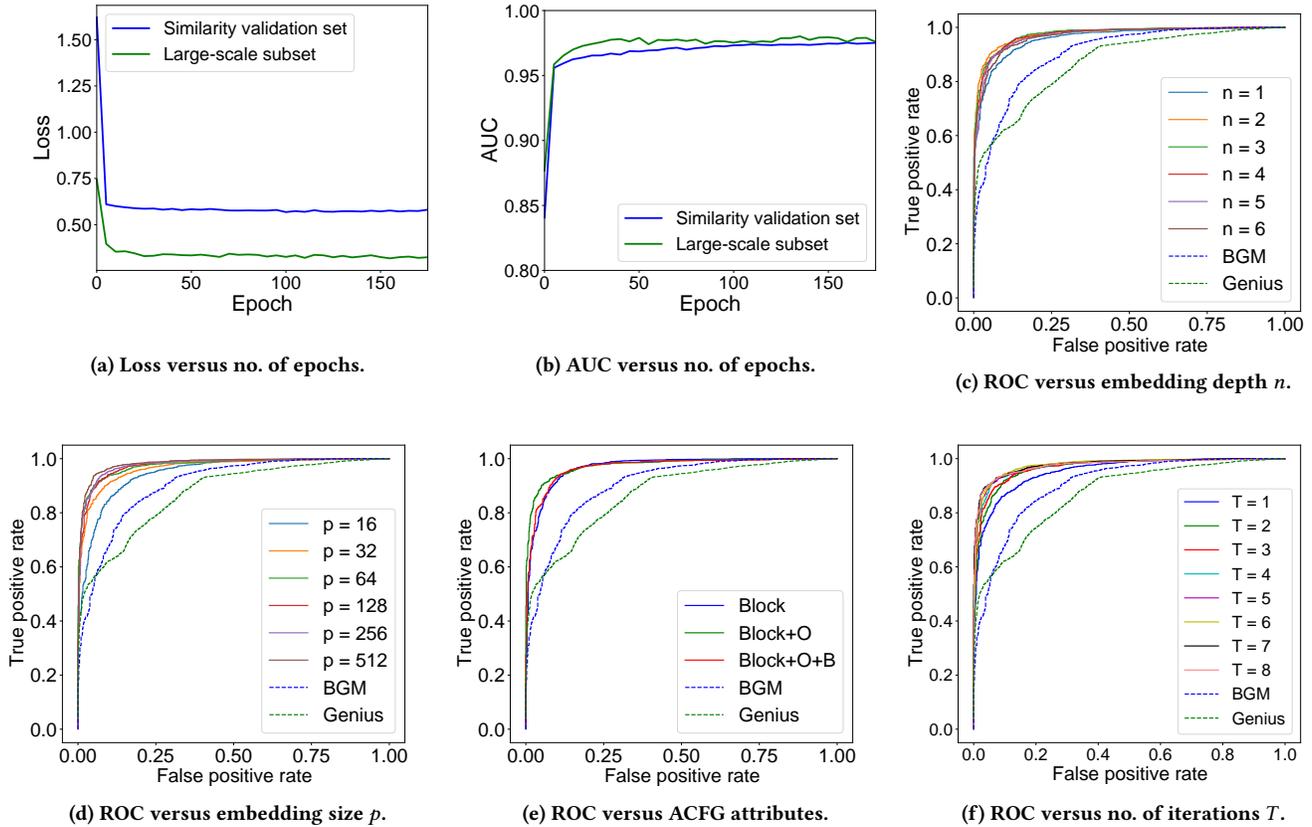

Figure 7: Effectiveness of different hyperparameters of NN. Figure 7a and Figure 7b are evaluated over the similarity validation set. Figure 7c, 7d, 7e, and 7f are evaluated over large-graph subset of the similarity testing set.

We further examine the performance of Gemini on GPU. However, we observe that the GPU version is on average 10% slower than the CPU version. This happens mostly on embedding generation for small graphs. On larger graphs, the GPU version can run faster than the CPU version by up to 70%. We attribute this observation to the fact that the GPU version requires additional overhead, i.e., allocating GPU memory and copying data from the main memory to the GPU memory, before the computation. Therefore, as the ACFG becomes larger, this overhead becomes insignificant when compared with the overall time for computing the embedding.

**Overall latency of embedding generation.** The embedding generation time for Genius includes ACFG extraction time for both block-level and inter-block features (i.e., Block+O+B in Figure 6a) as well as the multi-threaded CPU implementation of feature encoding (i.e., Genius (M) in Figure 6c). For Gemini, the embedding generation time includes ACFG extraction time for block-level features and the number of offspring (i.e., Block+O in Figure 6a) as well as the CPU implementation of graph embedding (i.e., Gemini (CPU) in Figure 6c). We observe that Gemini can achieve a speedup ranging from 27.7× to 11625.5×. On average, Gemini runs 386.4× faster than Genius.

### 4.5 Training time

Although offline training time will be amortized over a large number of online queries, the release of new firmware images may require the learning model to be updated on a monthly-basis or even weekly-basis to more accurately model the data. Therefore, we briefly compare the training time of Genius and Gemini.

The Genius approach needs to compute a codebook using an unsupervised learning algorithm called spectral clustering. This algorithm requires constructing a distance matrix which requires quadratic time complexity on the size of the training data. As a result, when the training data contains 100,000 functions (which is used to construct the codebook in our previous experiments), Genius takes more than one week to construct the codebook.

In contrast, since Gemini model runs only a fixed number of epochs, its running time is linear to the number of epochs and also linear to the number of samples in each epoch, i.e., the training dataset. In our experiment, each epoch contains around 206,000 training samples and takes less than 5 minutes to run. Earlier we have shown that the Gemini model requires only 5 epochs of training to achieve a reasonable performance, while running for 100 epochs gives the best performance. That means, Gemini requires less than 30 minutes (5 epochs) to train a model to outperform Genius, while achieving the best performance requires less than 10 hours. Therefore, our neural network-based approach allows the model to be updated more frequently than Genius. This property is crucial to make retraining and model update practical.

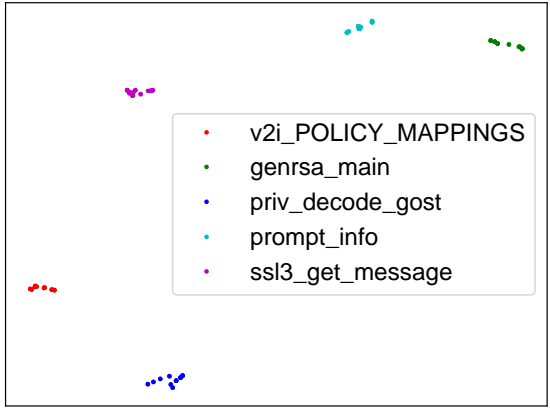

Figure 8: Visualizing the embeddings of the different functions using t-SNE. Each color indicates one source functions. The legend provides the source function names.

### 4.6 Understanding the Embeddings

We visualize the embeddings computed using the task-independently pre-trained embedding network to understand its effectiveness. In particular, we randomly select 5 source functions, and compute the embeddings of the corresponding binary functions compiled using different compilers, different target architectures, and different optimization-levels. We then use t-SNE [46] to project the high-dimensional embeddings onto a 2-D plane. We plot the projected points in Figure 8, and different source functions are presented in different colors. We can observe that (1) binary functions compiled from the same source function are close to each other; and (2) binary functions compiled from different source functions are far from each other. Therefore, this visualization illustrates that our task-independent pre-trained embedding function can preserve the information of the source function in the embeddings regardless of the target architectures, the compilers being used, and the optimization levels.

### 4.7 Accuracy of Task-specific Retrained Model using Real-world Dataset

In this section, we evaluate the effectiveness of our task-specific retraining approach using real-world firmware images. The evaluation is setup in the same way as in [18]. We extract the ACFGs of functions in Dataset II, which in total result in 420,558,702 functions. We further choose two vulnerabilities from Dataset IV, which are the same as used in [18]. For each vulnerability, we consider it as a *specific task* to search for as many functions as possible in Dataset I that contain the same vulnerability. To achieve this, we retrain the model from the one pre-trained on Dataset I.

To compare with [18], which inspects only the top-50 most similar results, we also evaluate the precision among the top-50 functions. We show that Gemini, through retraining, can achieve over 80% accuracy for each task among top 50 results, which is significantly better than prior art [18] whose accuracy is around 20% to 50%. We present the details below. In total, our approach allows identifying more than 25 novel vulnerable firmware images on average from than Genius among the top-50 results.

**Retraining effectiveness.** We perform retraining in an iterative manner. At the beginning, we have a pre-trained model, and use it to compute the embeddings of all functions in the target corpus to build an index. In this case, each query can be handled within 3 seconds. We manually inspect the top-$K$ (e.g., $K = 50$) results, and assign ground truth labels for each of them so that the top-$K$ results are used for retraining. After each iteration of retraining, we re-compute the embeddings for a subset (e.g., 10%) randomly sampled from the entire target corpus to get a new list of top-$K$ results. We repeat this process for several iterations. In practice, our experiments show that we only need very few iterations of retraining. Note that the Genius approach does not provide the flexibility to incorporate such additional supervision efficiently. Thus the retraining process is a unique advantage of our approach over Genius and allows our approach to achieve higher accuracy with additional supervision from domain experts.

In particular, we use the same two vulnerabilities as used in Genius, i.e., CVE-2015-1791 and CVE-2014-3508, since Genius [18] provides reference results. For CVE-2015-1791, we find that after only 1 iteration of retraining, our approach discovers 42 true positives from four vendors such as D-Link, ZyXEL, DD-wrt, and Tomato by Shibby, among top-50 results. Further, among top-100 results, our approach discovers 85 true positives. These results show that one retraining iteration helps improve the precision to over 84%. As a reference, Genius can detect only 14 vulnerable firmware images (a precision at 28%), and they are only from D-Link and Belkin.

CVE-2014-3508 is harder, since the control flow graphs do not change before or after the patch, but only one more instruction, i.e., storing a zero value into the memory buffer, is inserted as the patch. Using our approach, after 3 iterations of retraining, our approach identifies 41 firmware images among the top-50 results that are vulnerable, which results in a 82% precision. In comparison, Genius can identify only 24 vulnerable firmware images (a precision at 48%) among the top-50 results.

**Retraining time.** In terms of time consumption, for each iteration, we retrain the model for 5 epochs, and sample 10% of the entire dataset for evaluation (as discussed earlier). Notice that in the second step, the ACFGs did not need to be regenerated, and thus we only need to pay the cost for embedding computation. The overall time of these two automated steps can be done within 2 hours for each iteration. In terms of manual investigation time, we find that an experienced expert can finish the manual investigation of 50 candidates within 2 hours. This time can be even shorter for later iterations, since the vulnerable code is already familiar to the experts after the first iteration of investigation. In total, a human expert spend less than 12 hours in going through three iterations to train an effective model for a given vulnerability. Once the model is deployed to the entire dataset, it took around 12 hours to generate the embeddings for the entire dataset.

Therefore, we conclude that the retraining capability of our approach enables practical usage of feedback from human experts to increase the search accuracy within a reasonable amount of time, i.e., within one day.

## 5 RELATED WORK

We have discussed closely related work throughout the paper. In this section, we briefly survey additional related work. We focus on approaches using code similarity for known bugs search without source code. Other approaches for finding unknown bugs [3, 8, 9, 34, 42–44] will not be discussed in this section. For learning-based bug search [18], we have already discussed the comparison earlier in the paper.

**Raw feature based bug search.** Many researchers have already worked on the problem of bug search in binaries, and made great contribution towards this direction. Fundamentally, they rely on various raw features directly extracted from binary for code similarity matching. N-grams or N-perms [26] are two early approaches for the bug search. They adopt the binary sequence or mnemonic code matching without understanding the semantics of code [25], so they cannot tolerate the opcode reordering issue caused by different compilations. To further improve the accuracy, the tracelet-based approach [14] captures execution sequences as features for code similarity checking, which can address the opcode changes issue. TEDEM [32] captures semantics using the expression tree for each basic block. However, the opcode and register names are different across architectures, so these two approaches are not suitable for finding bugs cross architectures.

Many other approaches can be used for bug search in the cross-architecture setting, but they are expensive to be applied for large scale firmware bug search. Zynamics BinDiff[15] and BinSlayer [6] adopt the expensive graph isomorphism algorithm to quantify the similarity between control flow graphs for code search. The symbolic execution and theorem prover used in BinHunt[20] and iBinHunt [29] are by design expensive, since they need to extract the equations and conduct the equivalent checking.

Although Pewny et al. [31] use MinHash to reduce code similarity computation, their graph matching algorithm is still too expensive to handle millions of graph pairs. DiscovRE [16] utilizes the pre-filtering to boost CFG based matching process by eliminating unnecessary matching pairs, but the pre-filtering is unreliable and outputs tremendous false negatives [18]. Many other approaches, such as Costin et al. [11], are indeed efficient when searching bugs at large scale, but they are only designed for specific bugs with obvious artifacts, and cannot handle more general cases.

**Graph embedding.** Graph analysis has its significance in various real-world application, such as biology [13] and social network [19]. Typically, there are two different meaning of graph embedding that are used in graph analysis. The first one is to embed the nodes of a graph. This means finding a map from the nodes to a vector space, so that the structural information of the graph is preserved [21]. Among early approaches, LLE[36] finds the embedding vectors so that the embedding of a node is a linear combination of the nodes near it. In [4], the embedding of two nodes are close to each other when the weight of the edge between them is large. Recently, deep-learning-based method is adopted to deal with large scale graph dataset.

Another meaning of graph embedding, which is adopted in this paper, is to find a embedding vector that represents the whole graph. After that, people can perform machine learning methods on it to deal with tasks like protein design and gene analysis [39]. Currently, kernel method [38] is a widely used technique for processing structural data like sequences[17] and graphs[5].

The key to the kernel method is a carefully designed kernel function (a positive semidefinite function between pairs of nodes). A class of kernels are designed by counting the elementary structures that appears in the graph. For example, [33] counts specific subtree patterns in a graph; [41] counts the appearance of subgraph with specific sizes; in [40], different structures will be counted in a process named Weisfeiler-Lehman algorithm. However, in these methods the kernels are fixed before learning, so the embedding space may have very large dimensions.

Another class of kernels leverages the fact that graphical models can take into account structured data with noise and variations. Two representative examples are the Fisher kernel [23] and the probability product kernel [24]. These kernels fit graphical models for input graphs and use some form of inner product between the distributions as the kernel function. The model applied in our paper, `Structure2vec`[13], also construct graphical models for input graphs, and parameterize the inference algorithms of graphical models by neural network to define features for the corresponding kernel function.

**Deep learning-based graph embedding approaches.** Scarselli et al. proposed the first graph neural network to compute embeddings of a graph [37]. Li et al. extend [37] by using Gated Recurrent Unit (GRU) to generate features. Dai et al. generalize both works using principled graphical model thinking which allows more flexible embedding functions to be defined [13]. Therefore, we use a variant of [13] as our embedding generation function.

There is another line of research to generate graph embeddings from large networks such as social-networks [22, 30, 45, 47]. These works focus on unsupervised learning or semi-supervised learning and generating features of different nodes in a graph rather than the embeddings of the entire graph. Using these approaches, it is also not easy to incorporate the additional training data into retraining, and thus they are not suitable for our problem.

## 6 CONCLUSION

In this paper, we present a deep neural network-based approach to generate embeddings for binary functions. We implement a prototype called Gemini. Our extensive evaluation shows that Gemini outperforms the the state-of-the-art approaches by large margins with respect to similarity detection accuracy, embedding generation time, and overall training time. Our real world case studies demonstrate that using retraining Gemini can identify significantly more vulnerable firmware images than the state-of-the-art, i.e., Genius.

Our research showcases a successful application of deep learning on computer security problems.

## ACKNOWLEDGEMENT

We thank the anonymous reviewers for the helpful comments. We thank Xinyun Chen for her help to write this paper. This material is in part based upon work supported by the National Science Foundation under Grant No. TWC-1409915, 1664315, 1719175, IIS-1350983, IIS-1639792, and SaTC-1704701, ONR under N00014-15-1-2340, DARPA under FA8750-15-2-0104 and FA8750-16-C-0044, Berkeley Deep Drive, NVIDIA, Intel and Amazon AWS. Any opinions, findings, and conclusions or recommendations expressed in this material are those of the author(s) and do not necessarily reflect the views of the National Science Foundation.